\documentclass[12pt]{article}
\newcommand{\be}{\begin{equation}}
\newcommand{\ee}{\end{equation}}
\newcommand{\bea}{\begin{eqnarray}}
\newcommand{\eea}{\end{eqnarray}}
\newcommand{\ba}{\begin{array}}

\newcommand{\ea}{\end{array}}

\def\bbox{{\,
\lower0.9pt\vbox{\hrule \hbox{\vrule height 0.2 cm
\hskip 0.2 cm \vrule height 0.2 cm}\hrule}\,}}
\newcommand{\dsl}{\pa \kern-0.5em /}

\newcommand{\nn}{\nonumber \\}
\newcommand{\bra}{\langle}
\newcommand{\ket}{\rangle}
\newcommand{\sgn}{{\rm sgn\,}}
\newcommand{\sech}{{\rm sech\,}}

\def\tr{{\rm tr\,}}
\def\Tr{{\rm Tr\,}}

\def\CL{{\cal L}}                       

\def\CH{{\cal H}}

\def\CV{{\cal V}}

\def\ds{\raise.15ex\hbox{/}\kern-.57em\partial}
\def\Ds{\,\raise.15ex\hbox{/}\mkern-13.5mu D}
\begin{document}

\baselineskip 18pt


\begin{titlepage}
\vfill
\begin{flushright}
SNUTP01-036\\
hep-th/0110154\\
\end{flushright}

\vfill

\begin{center}
\baselineskip=16pt
{\Large\bf One-loop Effective Actions in Shape-invariant Scalar Backgrounds}
\vskip 10mm
{Chanju Kim,$^{1}$ O-Kab Kwon,$^{2}$ and Choonkyu Lee$^{3}$}
\vskip 8mm
{\small\it
Department of Physics and Center for Theoretical Physics\\
Seoul National University, Seoul, 151-747,
Korea \\}
\vskip 10 mm
\par
{\bf ABSTRACT}
\begin{quote}
The field-theoretic one-loop effective action in a static scalar background
depending nontrivially on a single spatial coordinate is related, in the
proper-time formalism, to the trace of the evolution kernel (or heat kernel)
for an appropriate, one dimensional, quantum-mechanical Hamiltonian. We
describe a recursive procedure applicable to these traces for shape-invariant
Hamiltonians, resolving subtleties from the continuum mode contributions by
utilizing the expression for the regularized Witten index. For some cases 
which include those of domain-wall-type scalar backgrounds, our recursive 
procedure yields the full expression for the scalar or fermion one-loop 
effective action in both (1+1) and (3+1)-dimensions.
\end{quote}
\end{center}

\vfill
\vskip 5mm
\hrule width 5.cm
\vskip 5mm
\begin{quote}
{\small
\noindent
$^1$ E-mail: cjkim@phya.snu.ac.kr\\
$^2$ E-mail: kok@phya.snu.ac.kr\\
$^3$ E-mail: cklee@phya.snu.ac.kr\\
}
\end{quote}
\end{titlepage}
\setcounter{equation}{0}

The calculation of the one-loop effective action or one-loop quantum
correction to the vacuum energy in a static background configuration is an
important problem in renormalizable field theory which received much attention
over the years \cite{Schwinger}-\cite{DT}.
If the background classical configuration has a
nontrivial spatial dependence as in the case of a soliton background, it
becomes a rather difficult task. This is because one must evaluate certain
nontrivial functional determinant to determine the effect due to fluctuations
of quantum fields in the presence of such a background configuration.
Renormalization aspect is now well understood, however.

In this paper we present some interesting observations which are relevant for
the scalar or fermion one-loop effective action in a static scalar background
having nontrivial dependence on a single spatial coordinate only.
Specifically, we shall employ the Schwinger proper-time representation
\cite{Schwinger}.
Then the principal task in determining the one-loop effective action
reduces to the calculation of the trace of the evolution kernel (or heat
kernel) for a suitable, one-dimensional, quantum-mechanical Hamiltonian. We
are here interested in a family of background scalar fields which lead, in
this proper-time formalism, to so-called shape-invariant Hamiltonians
\cite{GP} (as formulated on the basis of supersymmetric quantum
mechanics \cite{Witten}). For
a family  of shape invariant Hamiltonians, we will below derive the recursive
relation applicable to the corresponding traces; this may in turn be used to
relate the one-loop effective actions for different scalar backgrounds in the
given family. As will be considered in the latter portion of this paper,
applying this scheme to the one-loop energies of kink configurations produces
certain results of direct physical interest.

First let us consider a system involving a quantized
Dirac field in the presence of an external
scalar background $\tilde\phi(x)$, described by the Lagrangian
\be
\CL = \bar\psi(x) ( i \gamma^\mu \partial_\mu  + \phi(x))\psi(x),
\quad (\phi \equiv -m + \tilde\phi(x)).
\ee
The (unrenormalized) effective action $W_0(\phi)$ may then defined through the
path integral
\be\label{wdef}
e^{iW_0(\phi)} = N \int[D\psi][D\bar\psi]\exp\left(i\int d^dx \CL \right),
\ee
where $d$ is the space-time dimension and $N$ the normalization factor chosen
such that $W_0(\phi)$ may vanish if there is no background field for instance.
In the Schwinger proper-time representation
\cite{Schwinger}, it can be expressed by the form
\be \label{w0}
W_0(\phi) = \frac{i}2 \int_0^\infty \frac{ds}s e^{-s\epsilon}
            \int d^dx\, \tr  \bra xs| x \ket ,
\ee
up to a suitable, normalization-related, additive constant.
Here, $\bra xs | x \ket$ is the coincidence limit of the proper-time
Green's function,
\be
\bra xs| x' \ket = \bra x | e^{-is \CH} |x' \ket,
\ee
for the {\em squared} Dirac operator $\CH$ which has the explicit form
\be
\CH = -\partial^2 + \phi^2 + i \gamma^\mu \partial_\mu \phi.
\ee

We are here interested in the case when the background field $\tilde\phi$
depends only on a single spatial coordinate which is denoted as $z$.
Then the above `Hamiltonian' $\CH$ has an interesting structure: aside from
the trivial transverse derivative terms (including the time derivative),
$\CH$ defines a one-dimensional
supersymmetric quantum mechanics \cite{CKM} with superpotential $\phi$.
In the basis where $\gamma^\mu \partial_\mu z$ is given by a diagonal
matrix, $\CH$ becomes
\bea
\CH &=& -\partial_\perp^2 + \pmatrix{ - \partial_z^2 + \phi^2 + \phi' & 0 \cr
                                  0   & - \partial_z^2 + \phi^2 - \phi' } \nn
    &=& -\partial_\perp^2 + \pmatrix{ \CH_+ & 0 \cr
                                      0   & \CH_-}
\eea
where $\phi' \equiv d\phi/dz$ and
$\partial_\perp^2 \equiv \partial^2 - \partial_z^2$.
The one-dimensional Hamiltonians
$\CH_\pm$ form a supersymmetric pair and can be written as
\be \label{superh}
\CH_+ = QQ^\dagger, \quad
\CH_-= Q^\dagger Q,
\ee
where
\be \label{supercharge}
Q = -i\frac{d}{dz} - i \phi, \quad
Q^\dagger = -i\frac{d}{dz}+ i \phi.
\ee
We also remark that, in a {\em static\/} background field, the effective
action per unit time interval coincides, up to the sign, with the one-loop
energy of the system defined by the background configuration \cite{Rajaraman}.

In order to obtain the effective action $W_0$ in (\ref{w0}), we need
to evaluate the quantity
\be
\Tr e^{-is\CH_+} + \Tr e^{-is\CH_-},
\ee
where we defined $\Tr \equiv \int dz\, \tr$.
An important property of the supersymmetric system is that the two
Hamiltonians $\CH_\pm$ are almost
isospectral, that is, they share all the eigenvalues except for the zero
eigenvalue of, say, $\CH_-$. This enables us to relate one of the
above traces
with the other through the celebrated (regularized) Witten index
$\Delta(is)$ \cite{Witten2},
\be \label{wittenindex}
\Delta(is) = \Tr e^{-is\CH_-} - \Tr e^{-is\CH_+}.
\ee
Essentially $\Delta(is)$ counts the difference between the number of zero
modes of $\CH_-$ and $\CH_+$.
There are, however, some subtleties from the continuum mode
contributions, which can make the index fractional \cite{AC}.
Nevertheless, it is known that the regularized Witten index is a topological
invariant, i.e., is invariant under any compact perturbations of the
background field. (For a particularly simple proof of this fact, see
\cite{BB}.) Hence it can be
calculated largely independently of the details of the Hamiltonian.
Indeed, for any background field $\phi(z)$ which
smoothly interpolate between two nonzero
values $\phi_\pm$ at $z = \pm\infty$, the Witten index is given purely
in terms of $\phi_\pm$ \cite{AC},
\be \label{wittenindex2}
\Delta(is) = \frac{\sgn(\phi_+)}2 \left[
                1- \frac1{\sqrt\pi}\Gamma(1/2,is\phi_+^2) \right]
            - \frac{\sgn(\phi_-)}2 \left[
                1- \frac1{\sqrt\pi}\Gamma(1/2,is\phi_-^2) \right],
\ee
where $\Gamma(a,x)$ is the incomplete gamma function.
Therefore we need only to evaluate one of the traces, say, $\Tr e^{-is\CH_+}$.

Similar consideration can also be given to the effective action for a
quantized scalar field which is described by the Lagrangian
\be \label{lags}
\CL = -\frac12 \partial_\mu\varphi \partial^\mu\varphi
    -U(\varphi).
\ee
The one-loop effective action due to quantum fluctuations of the scalar field
in the presence of the static background field $\phi(z)$ becomes
\be \label{w0s}
W_0(\phi) = -\frac{i}2\int_0^\infty \frac{ds}s e^{-s\epsilon}
            \int d^dx  \bra x|e^{-is\CH}| x \ket ,
\ee
where
\bea \label{scalarh}
\CH &=& -\partial_\perp^2 - \partial_z^2
        + U''(\phi(z)) \nn
    &\equiv& -\partial_\perp^2 + \CH_s.
\eea
Clearly, we are led to consider a quantum mechanical Hamiltonian
analogous to that of the Dirac field case.

So far we have seen that, in determining the one-loop effective action in a
static scalar background, it becomes necessary
to calculate the trace of the evolution
kernel for a certain quantum mechanical Hamiltonian. In general,
such a task cannot be performed exactly and one has to rely on certain
approximation methods such as the WKB method. For a certain class of
scalar background
$\phi(z)$, however, the resulting quantum mechanical system is exactly
solvable and it has been known that such a system usually possesses a special
`symmetry' known as shape invariance \cite{GP}. We shall call such a scalar
background simply as a shape-invariant scalar background. In what follows,
our direct concern will be on the trace of the evolution kernel in such
shape-invariant scalar background.

Let $c_0$ be a parameter appearing in the background scalar field $\phi(z)$,
and then we have the Hamiltonians $\CH_\pm(c_0)$ defined by (\ref{superh}) 
and (\ref{supercharge}), Here,
$\CH_+(c_0)$ is called shape invariant \cite{GP} when
\bea
\CH_+(c_0) &=& Q(c_0)Q^\dagger(c_0) \nn
           &=& Q^\dagger(c_1)Q(c_1) + R_1 = \CH_-(c_1) + R_1,
\eea
where $c_1$ is a new parameter depending on $c_0$ and $R_1$ is another 
constant.  Now we can construct a sequence of Hamiltonians 
$\CH^{(n)}$ such that
\bea \label{hsequence}
\CH^{(0)} &=& \CH_-(c_0) \nn
\CH^{(1)} &=& \CH_+(c_0) = \CH_-(c_1) + R_1\nn
\cdots&& \nn
\CH^{(n)} &=& \CH_+(c_{n-1}) + \sum_{k=1}^{n-1} R_k \nn
          &=& \CH_-(c_n) + \sum_{k=1}^{n} R_k .
\eea
Note that each pair $\CH_+(c_k)$ and $\CH_-(c_k)$ itself defines a new
supersymmetric quantum mechanical system, and the corresponding Witten
index
\be
\Delta_k(is) = \Tr e^{-is\CH_-(c_k)} - \Tr e^{-is\CH_+(c_k)}
\ee
may be calculated separately without much effort using (\ref{wittenindex2}).
Then (\ref{hsequence}) allows us
to relate the trace of the evolution kernel for $\CH_+(c_0)$ to that for
$\CH_+(c_n)$ as
\bea \label{recursion}
\Tr e^{-is\CH_+(c_0)} &=& e^{-isR_1} \Tr e^{-is\CH_-(c_1)}\nn
     &=& e^{-isR_1} \left[ \Tr e^{-is\CH_+(c_1)} + \Delta_1(is)\right]\nn
     &=& \cdots\nn
     &=& e^{-isE_n} \Tr e^{-is\CH_+(c_n)}
           + \sum_{k=1}^n e^{-isE_k} \Delta_k(is),
\eea
where $E_k = \sum_{i=1}^{k} R_i.$
Therefore, once the trace is calculated for one shape-invariant Hamiltonian,
the traces for the rest of Hamiltonians in the same shape-invariant family
can be obtained immediately from (\ref{recursion}) without new calculations.
This is especially useful when for some value of $n$ the
Hamiltonian $\CH_+(c_n)$ reduces to a simple one. It will then allow us to
calculate the effective action (\ref{w0}) almost effortlessly for certain
nontrivial scalar backgrounds.

Similar development can be given for the trace defined by the scalar
Hamiltonian $\CH_s$, if the potential term $U''(\phi)$
produces a
shape-invariant potential. Of course, the background field $\phi$ is
no longer identified as the superpotential appearing in the operator $Q$; 
but, with a suitable superpotential, say, $\bar\phi$, the Hamiltonian may 
still be written in a shape-invariant form. This is the case when the 
background field $\phi$ corresponds to the kink solution of the 
scalar theory.

With the trace of the evolution kernel at hand, one still has to perform the
integration over the proper time $s$ to obtain the effective action 
via (\ref{w0}) or (\ref{w0s}).
In doing so, ultraviolet divergences appear from
the $s \rightarrow 0$ end --- the expression should be
renormalized. Here one may utilize the Dewitt
WKB expansion \cite{DeWitt}
\be \label{wkb}
s \rightarrow 0 : \bra x|e^{-is\CH}|y \ket
= \frac{i}{(4\pi is)^{d/2}}e^{i(x-y)^2/4s - im^2s}
    \sum_{k=0}^\infty a_k(x,y) (is)^k,
\ee
when $m$ is the mass parameter of the corresponding free theory. 
The coefficient functions
$a_k(x,y)$ are well-behaved in the region including the coincidence point
$x=y$ and can be determined using the Schr\"odinger-type equation satisfied
by the evolution kernel. For
\be
\CH = -\partial^2 + m^2 + V,
\ee
the coincidence limits of first few coefficient functions are given by
\cite{Ball}
\be \label{ai}
a_0(x,x) = 1, \quad
a_1(x,x) = -V, \quad
a_2(x,x) = -\frac{1}{6} \partial^2V + \frac12 V^2.
\ee
It is now clear from (\ref{w0}) or (\ref{w0s}) that all the divergences
result from the terms involving the coincidence limits $a_0(x,x), \ldots,
a_{[d/2]}(x,x)$ in the series for $\bra xs|x \ket$, and 
the renormalized effective action can be obtained if those terms are 
subtracted away\footnote{Here
the dimensional regularization is implicitly assumed. In general, the precise
forms of the subtraction terms will depend on the renormalization
scheme used.}.

As an illustration, we calculate the fermion effective action in
(1+1) dimensions when the background field $\phi(z)$ is given by the form
\be \label{phi}
\phi(z) = c_0 \tanh{z}.
\ee
This is a typical situation for a fermion having the Yukawa-type interaction
with the scalar field corresponding to the kink configuration. (See
(\ref{kink}) below.) 
It is easy to see that it generates a family of
shape-invariant Hamiltonians $\CH_\pm^{(k)}$ with superpotentials
\be \label{phik}
\phi^{(k)} =c_k \tanh{z}, \quad c_k = c_0 - k.
\ee
The constant $R_k$ defined in (\ref{hsequence}) is given by
\be
R_k = c_{k-1}^2 - c_k^2 = 2(c_0 - k) + 1.
\ee
Now, for systems defined by the superpotentials in (\ref{phik}), we may use
the formula (\ref{recursion}) in conjunction with
(\ref{wittenindex}) to get the relationship between the traces of the
corresponding evolution kernels. 
In particular, if the constant $c_0$ appearing in $\phi(z)$ is a positive 
integer, i.e.,
\be
c_0 = n,
\ee
the $n$-th superpotential $\phi^{(n)}$ vanishes and the corresponding
Hamiltonian becomes a {\em free} Hamiltonian. In such situation,
one can readily obtain the effective action in the given nontrivial 
background by using our relationship. Explicitly, from (\ref{w0}), the
unrenormalized effective action is then given by
\bea \label{w0result}
W_0 &=& \frac{i}{2^{2-d/2}} \int_0^\infty \frac{ds}{s}
     \int d^{d-1}x_\perp \bra x_\perp | e^{is\partial_\perp^2} |x_\perp \ket\,
     \Tr( e^{-is \CH_-} + e^{-is\CH_+} ) \nn
    &=& \frac{i \CV_\perp}{2^{2-d/2}} \frac{i}{(4\pi i)^{(d-1)/2}}
      \int_0^\infty \frac{ds}{s^{(d+1)/2}} \left[
      2 e^{-is E_n}\, \Tr e^{is \partial_z^2}
     + 2 \sum_{k=1}^n e^{-is E_k} \Delta_k(is) + \Delta_0(is) \right],\nn
&&
\eea
where $\CV_\perp$ is the space-time volume of the transverse directions
and $E_k$ is given by
\be
E_k = 2nk-k^2.
\ee
The Witten index $\Delta_k(is)$ is calculated using the formula
(\ref{wittenindex2}) as
\be
\Delta_k(is) = 1 - \frac{1}{\sqrt\pi} \Gamma(1/2, is(n-k)^2).
\ee
The first term in the integrand of (\ref{w0result}) produces the effective
action appropriate to the free theory with mass $m = n (=c_0)$. It may be 
cancelled by
the normalization factor $N$ in (\ref{wdef}) and we subtract it
from the effective action. Even after such subtraction, the remaining 
expression is still divergent; to secure a finite expression, 
further renormalization must be considered.

Let us first consider the
(1+1) dimensional case. We here use the procedure described already, i.e.,
apply the Dewitt WKB expansion (\ref{wkb}) to the squared Dirac operator
with $\phi(z)$ in (\ref{phi}), writing
\be
-\partial_\perp^2 + \CH_\pm = -\partial^2 + m^2 + V_\pm,
\ee
where $m^2 = n^2 (= c_0^2)$ is the free particle mass and
\be
V_\pm = \phi^2 \pm \phi' - m^2.
\ee
Now the divergent terms are cancelled by the subtraction of the 
$a_1^\pm(x,x)$ term in (\ref{ai}), that is, by considering the counterterm
\be \label{deltaw}
\delta W_c = \frac{i}2\int_0^\infty \frac{ds}s e^{-s\epsilon} \int d^dx
   \bra x | e^{-is(-\partial^2 +m^2)} |x \ket is[a_1^+(x,x) + a_1^-(x,x)],
\ee
with
\bea
a_1^+(x,x) + a_1^-(x,x)
         &=& -( 2\phi^2 -2 m^2) \nn
         &=& 2 n^2 \sech^2 z.
\eea
Then the renormalized effective action is given by
\bea
W &=& -\frac{T}{2(4\pi i)^{1/2}}
      \int_0^\infty \frac{ds}{s^{3/2}} \left\{
      2 \sum_{k=1}^n e^{-is(2nk-k^2)}\left[
         1 - \frac{1}{\sqrt\pi} \Gamma(\frac12, is(n-k)^2) \right]\right.\nn
 && \left. \hspace{36mm}
      - \frac{1}{\sqrt\pi} \Gamma(\frac12, isn^2)
      - \frac{e^{-isn^2}}{\sqrt{4\pi is}} \int dz\; 2isn^2\sech^2 z\right\},
\eea
where $T=\CV_\perp \equiv \int dt$ and we have ignored the zero mode 
contribution in $\Delta_0(is)$ as usual \cite{Rajaraman}. 
The integral over $s$ can be performed explicitly and we obtain
\be \label{wrf}
W = -T \left[ \frac{2}{\pi}
  \sum_{k=1}^{n-1} \sqrt{n^2-k^2} \tan^{-1}\left(\frac{\sqrt{n^2-k^2}}k\right)
  - \sum_{k=1}^{n-1} \sqrt{n^2-k^2} + \frac{n^2}{\pi} \right].
\ee
The result is in exact agreement with the one-loop energy correction
calculated in \cite{CY}.

In (3+1) dimensions, on the other hand, we need to subtract terms involving
the coefficients $a_2(x,x)$ also. The counterterm is thus given by
\bea
\delta W_c &=& \frac{i}2\int_0^\infty \frac{ds}s e^{-s\epsilon} \int d^dx
   \bra x | e^{-is(-\partial^2 +m^2)} |x \ket \nn
         & & \hspace{30mm} \times \left\{
   is[a_1^+(x,x) + a_1^-(x,x)] + (is)^2[a_2^+(x,x) + a_2^-(x,x)]\right\},\nn
          &&
\eea
and, for $\phi(z) = n \tanh z$, we have (from (\ref{ai}))
\be
a_2^+(x,x) + a_2^-(x,x)
         = \frac{n^2}3(2\cosh{2z} + 3n^2-1) \sech^4{z}.
\ee
Based on this information, we then obtain the renormalized effective action
\bea
W &=& W_0 - \delta W_c \nn
  &=& -\frac{\CV_\perp}{3\pi^2} \left[
  -2 \sum_{k=1}^{n-1} (n^2-k^2)^{\frac32}
                      \tan^{-1}\left(\frac{\sqrt{n^2-k^2}}k\right)
  +\pi \sum_{k=1}^{n-1} (n^2-k^2)^{\frac32} - \frac{1}{3} n^2(n^2-2) 
  \right]. \nn
&&
\eea

Next, we consider the spontaneously broken $\varphi^4$-type scalar field
theory with the potential energy function
\be \label{lags2}
U(\varphi) = \frac{\lambda}{4}(\varphi^2 - m^2/\lambda)^2,
\ee
and calculate the one-loop effective action in the background of a kink
solution
\be \label{kink}
\phi(z) = \frac{m}{\sqrt\lambda} \tanh\frac{m}{\sqrt2}z.
\ee
Rescaling the coordinate $z \rightarrow \sqrt{2}z/m$, we find that the 
one-dimensional quantum mechanical Hamiltonian $\CH_s$
in (\ref{scalarh}) is given by
\be
\CH_s = -\partial_z^2 + 6\tanh^2z - 2.
\ee
This corresponds to the shape invariant Hamiltonian with the superpotential
$\bar\phi = 2 \tanh{z}$ (and so $\bar\phi^{(2)} = 0$. See (\ref{phik}).
In other words, this is actually a special case
of the previous Dirac system and we can immediately obtain the desired
renormalized one-loop effective action in an entirely similar manner.
The result in (1+1) dimensions for the scalar one-loop effective action reads
\be 
W = Tm\left(\frac{3}{\sqrt{2}\pi} - \frac1{2\sqrt6}\right),
\ee
which is in agreement with the result in \cite{DHN}.
In (3+1) dimensions our procedure produces the following result:
\be
W = \CV_\perp \frac{\sqrt{3}m^3}{24\sqrt{2}\pi}.
\ee
The same result was also obtained in a recent paper \cite{Carvalho}.

Finally, we comment on the (0+1) dimensional (quantum mechanical) case where
the kink configuration corresponds to an instanton solution and the 
determinant factor is then related to the tunneling rate 
of the system \cite{Coleman}. 
The determinant has been calculated in \cite{JF} utilizing the shape
invariance and heat kernel in a somewhat similar fashion to the present work.
However, our method is much simpler since we have made more effective
use of supersymmetry, especially, by utilizing the expression for the 
Witten index with which most of the calculations have been dispensed with.

In summary, we employed the Schwinger proper-time formalism to represent
the scalar or fermion one-loop effective action in a static scalar background.
It is related to the trace of the evolution kernel for an appropriate
quantum mechanical Hamiltonian. In the case that the scalar background 
results in a shape-invariant system, we derived a recursive relation 
applicable to the traces for shape-invariant Hamitonians. Most of the
complications in the calculation disappear if the expression for 
the regularized Witten index is used in a judicious way. We applied 
this scheme to the scalar or fermion system in a kink background, to obtain 
the corresponding
one-loop effective actions in (1+1) and (3+1) dimensions explicitly.
\section*{Acknowledgment}
This work was supported in part by the BK21 project of the Ministry of
Education, Korea, and also by Korea Research Foundation Grant 2001-015-DP0085
(C.L.).

\end{document}